\documentclass[a4paper]{article}
\usepackage{INTERSPEECH2020, multirow, multicol}

\title{Exploiting Multi-Modal Features From Pre-trained Networks for Alzheimer’s Dementia Recognition}
\name{Junghyun Koo$^1$, Jie Hwan Lee$^1$, Jaewoo Pyo$^2$, Yujin Jo$^3$, Kyogu Lee$^1$}

\address{
  $^1$Music \& Audio Research Group (MARG), Seoul National University\\
  $^2$Electrical and Computer Engineering, Seoul National University\\
  $^3$College of Liberal Studies, Seoul National University}
\email{\{dg22302, wiswisbus, jwpyo, tera-yujin, kglee\}@snu.ac.kr}

\begin{document}

\maketitle

\begin{abstract}


Collecting and accessing a large amount of medical data is very time-consuming and laborious, not only because it is difficult to find specific patients but also because it is required to resolve the confidentiality of a patient's medical records.
On the other hand, there are deep learning models, trained on easily collectible, large scale datasets such as Youtube or Wikipedia, offering useful representations.
It could therefore be very advantageous to utilize the features from these pre-trained networks for handling a small amount of data at hand.
In this work, we exploit various multi-modal features extracted from pre-trained networks to recognize Alzheimer's Dementia using a neural network, with a small dataset provided by the ADReSS Challenge at INTERSPEECH 2020. The challenge regards to discern patients suspicious of Alzheimer’s Dementia by providing acoustic and textual data.
With the multi-modal features, we modify a Convolutional Recurrent Neural Network based structure to perform classification and regression tasks simultaneously and is capable of computing conversations with variable lengths.
Our test results surpass baseline's accuracy by 18.75\%, and our validation result for the regression task shows the possibility of classifying 4 classes of cognitive impairment with an accuracy of 78.70\%.

\end{abstract}

\section{Introduction}
Collecting a sufficient amount of electronic health records is a challenging task with various factors \cite{adibuzzaman2017big, edinger2012barriers}. Due to this problem, researchers in the medical field are often provided with only a small amount of data given.
Owing to the fact that deep learning techniques perform better on large amounts of data, a number of studies using machine learning techniques have been conducted to solve specific medical problems, regarding a limited number of data \cite{chen2017deep, shaikhina2015machine}.
Dementia is also one of many medical symptoms facing this situation.

Dementia, a syndrome in which there is deterioration in cognitive function beyond what might be expected from normal ageing, is mostly affected by Alzheimer’s Disease \cite{Burns2009AlzheimersD}.
There were previous researches with various approaches to recognize Alzheimer's Dementia \cite{fraser2016linguistic, karlekar2018detecting, di2019enriching, jo2019deep}, which has shown excellent performance.
However, datasets used in these works were sufficient with quantity than the one used in this paper.

The ADReSS challenge \cite{bib:LuzHaiderEtAl20ADReSS} at INTERSPEECH 2020 hosts two tasks: Alzheimer’s Dementia (AD) classification and Mini Mental Status Examination (MMSE) regression, while providing a refined dataset. The dataset is equally balanced of AD and non-AD participants with the metadata of age and gender.
Each data is a conversation in which participants, in both audio and text modalities, spontaneously describes the picture given by the investigator.
Participants of the challenge are suggested to solve hosted tasks using only the given data, where the numbers of train and test data are 108 and 48, respectively.

For recognizing AD with small amounts of data, we determined it would be beneficial to use both acoustic and textual features. 
Furthermore, we leverage models pre-trained on large scale datasets as feature extractor to get better representation.
To this end, this paper focus on exploiting various multi-modal features, and design suitable network architecture.
We compare 3 and 4 different acoustic and textual features, respectively, and use the hand-crafted (HC) feature and part-of-speech (POS) tagging as additional inputs. The usage of POS and HC is influenced by previous research, which has approved that using these features gained from transcript can improve the performance \cite{di2019enriching}. The proposed network is a modified version of Convolutional Recurrent Neural Network (CRNN); capable of computing conversations with variable lengths, and implemented with methods to fit with a small amount of data. Also, the model is able to compute using the acoustic feature only, without any metadata, which can be efficient considering the real-world situation.
Our experimental results show using features of the pre-trained network leads to performance gain than that of raw, and regression results imply the potential of network classifying classes of cognitive impairment based on MMSE score.

\section{Multi-Modal Features}

This work compares 3 different acoustic and 4 different textual features.
To obtain a speech signal corresponding to each utterance in the transcription, alignment of the transcription and the signal is done by using \cite{schiel1999automatic, kisler2017multilingual}. Hence, the following multi-modal feature extraction in this section is applied to the aligned data.

\subsection{Acoustic Features}

\begin{itemize}

	\item \textit{openSMILE features}:	The openSMILE v2.3 toolkit \cite{eyben2010opensmile} provides multiple features from raw audio files.
From the toolkit, we use the ComParE feature \cite{eyben2013recent} and the eGeMAPS feature \cite{eyben2015geneva}. For ComParE feature, using one-way ANOVA, we select 393 features (p$\leq$0.05), out of 6,373 concerning the efficiency of model capacity. 



\item \textit{VGGish}:    We use VGGish \cite{hershey2017cnn}  which is trained with \textit{Audio Set} \cite{gemmeke2017audio} for audio classification. The feature is composed of 128 feature dimensions, where each feature is extracted from audio with a length of 960ms. To handle different lengths of utterance, we use the average value of the extracted VGGish features. 

\end{itemize}

\subsection{Textual Features}
\begin{itemize}
\item\textit{Pre-trained language model features}:    
We exploit transformer \cite{vaswani2017attention} based language models, GPT \cite{radford2018improving}, RoBERTa \cite{liu2019roberta}, and Transformer-XL \cite{dai2019transformer}.
Pre-trained on large corpora, these language models have shown the effectiveness to improve performance over a wide range of natural language processing tasks. Sentence representations are obtained by averaging word embeddings via \cite{Wolf2019HuggingFacesTS}. The specific settings for the language models are as follows, GPT: openai-gpt, RoBERTa: roberta-base, Transformer-XL: transfo-xl-wt103. The feature dimensions of GPT and RoBERTa is 768, and Transformer-XL dimensions of 1024. Besides the aforementioned features, we also use 300-dimensional GloVe vectors \cite{pennington2014glove}.

\item\textit{Hand-crafted features}: We integrate three categories, psycholinguistic, repetitiveness, and lexical complexity features, as HC features, which reflect the features of Alzheimer’s.
Psycholinguistic features and repetitiveness\footnote{https://github.com/vmasrani/dementia\_classifier} are that suggested by \cite{fraser2016linguistic}, and lexical complexity is the Lexical Complexity Analyzer for Academic Writing (LCA-AW)\footnote{ https://github.com/Maryam-Nasseri/LCA-AW-Lexical-Complexity-Analyzer-for-Academic-Writing}. 
These token-level HC features are aggregated to the conversational-level by only averaging participant's utterance.
We select and use 23 features whose p-value from one-way ANOVA is less than 0.05 amongst a total of 42 features.
\end{itemize}

\begin{figure}[t]
  \centering
  \includegraphics[width=\linewidth]{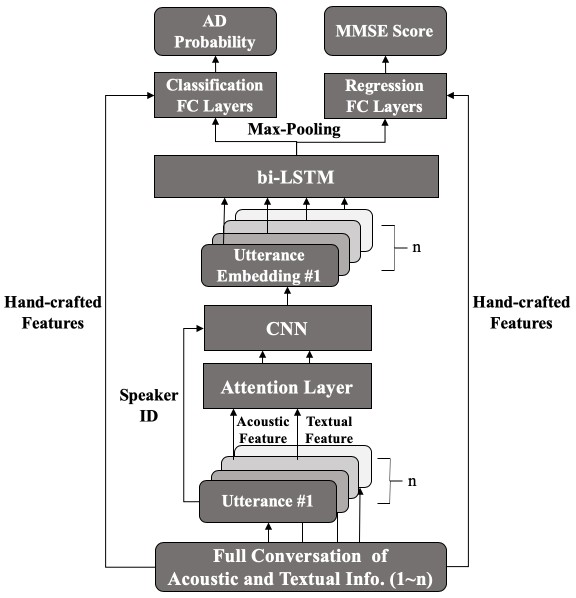}
  \setlength{\abovecaptionskip}{-5pt}
  \setlength{\belowcaptionskip}{-15pt}
  \caption{Overview of the proposed method.
  The acoustic and textual features  extracted from each utterance are fed in to the CRNN network. 
  Then, the hand-crafted features retrieved from the participant's entire conversation are concatenated to the utterance-level features.
  Finally, the FC layer of each task estimates the AD probability and MMSE score of the participant.}
  \label{fig:model_architecture}
\end{figure}

\section{Proposed Method}
While the proposed model can cope with additional inputs such as visual modality, the ADReSS challenge only offers acoustic and textual modalities. Thus, we primarily focus on the network with bimodal inputs. 
The overview of our model is as  Figure~\ref{fig:model_architecture}. In case of unimodal, the network has the same structure, except that only a single modality feature is input.

\subsection{Input}
An input dialogue consists of its utterances and an extracted HC feature. Each utterance comes along with an acoustic and a textual feature, and a speaker index. The speaker index is a binary feature denoting an investigator or a participant, where it is extended as the size of the largest size of input feature dimension, 1024 in our case, by a single fully connected layer. Input features smaller than 1024 are also expanded the same way by a fully connected layer.

We apply dropout \cite{srivastava2014dropout} to the input features before they are inserted into the network. This way, the model can be provided with more opportunity to learn independent representations, because each dimension can convey significant information, especially for the features extracted from pre-trained models.

\subsection{Model Architecture}
The proposed network is a modified version of CRNN, where an attention layer is a forefront layer of the network, and fully connected layers followed after the recurrent layer. Here, we use a bidirectional Long Short-Term Memory Network (bi-LSTM) \cite{huang2015bidirectional} as the recurrent network. 

Each modality input is individually inserted and computed through an attention layer. Our attention layer is implemented as the \textit{Scaled Dot-Product Attention} mechanism introduced in \cite{vaswani2017attention}. We use a self-attention mechanism, where an individual feature is used as a query, key, and value during the attentional computation.

Outputs of the attention layer and embedded speaker index of a single utterance are channel-wise concatenated then inserted into the one-dimensional Convolutional Neural Network (CNN). After a convolutional layer expands channel dimension to 32, 6 \textit{Squeeze-and-Excitation} (SE) \cite{hu2018squeeze} blocks are followed in the CNN. Each SE block consists of 2 convolution layers with a SE layer in between them. The last convolutional layer of every 2 SE blocks reduces feature dimension by convolutional stride factor of 4 and increments channel dimension. The expanding sizes of the channel dimension are 128, 512, 1024 respectively. Ultimately, CNN outputs 1024-dimensional channels with a global max pooled value.


After every utterance from the input dialogue is each computed through the CNN, the processed utterance embeddings are sequentially inputted into the bi-LSTM. The recurrent network consists of 3 bi-LSTM layers with 512 hidden units. Ultimately, the recurrent network outputs the max-pooled state from the results of the last layer’s hidden states and is concatenated with HC.

Three fully connected (FC) layers follow after the bi-LSTM layers. Both the first two FC layers are followed by a rectified linear unit (ReLU) activation and reduce the input dimension by a factor of 4. The last activation function for classification and regression tasks are softmax and sigmoid, respectively.
Ground truth MMSE score is scaled from 0 to 1 for regression loss computation.

\subsection{Training and Inference}
We use different numbers of utterances per batch during the training phase for the network to have opportunities to interpret various sequences of dialogue. The size is randomly selected between 5 and the minimum number of utterances among the dialogues in each batch. Since the minimum number of utterances of dialogue in the training data is 7, it was reasonable to set the minimum length to 5. If the length is too short, the network could be vulnerable to utterances with less meaningful data such as the investigator’s “okay” or “mhm”.
A single batch is used during the inference phase to analyze every utterance in an input dialogue.

Our training loss for classification and regression tasks are binary cross-entropy error and mean squared error, respectively. The total cost function is a summation of these two values. We use the Adam optimizer \cite{kingma2014adam} with a learning rate of 0.0002 and momentum parameters $\beta$1 = 0.5, $\beta$2 = 0.9.

\section{Experiments}
In this section, we evaluate model performances for both classification and regression tasks. Recorded performances are averaged value from measurements of 5-fold cross validation, where each fold contains 86 training and 22 validation conversations, except for the last fold containing 88 training and 20 validation conversations.

Prior experiments were conducted for optimizing several hyperparameters in the proposed network. First, we compared model performance with a one-dimensional convolutional kernel size of 3, 5, 10, 15. Through observations, larger kernel sizes led to performance gain; thus, we set the kernel size to 15. Attempts to ascertain the ideal dropout rate among 0 to 50\% at 10\% intervals could not be determined. Yet, we adopted a 20\% dropout rate for data augmentation and prevention of overfitting. Finally, we discovered using 6 instead of 3 stacked convolutional blocks achieved better performance. Experimental results of each model shown in this section share above achieved hyperparameter values.

\subsection{Feature Comparison}

\begin{table}[t]
  \caption{
Validation Results of Acoustic Unimodal Network
  }
  \label{tab:unimodal auditory experiment results}
  \centering
  \begin{tabular}{c|ccc}
    \hline
    Feature         & Accuracy          & F1               & RMSE   \\ \hline
    \textit{eGeMAPS} & 61.82\%           & 71.98\%          & 6.7178          \\
    \textit{ComParE} & 68.27\%           & 74.62\%          & 6.7852          \\
    \textit{VGGish}  & \textbf{85.27\%}  & \textbf{86.28\%} & \textbf{5.1144} \\ \hline
  \end{tabular}
\vspace{-4mm}
\end{table}

\begin{table}[t]
\caption{
Validation Results of Textual Unimodal Network
}
\label{tab:unimodal textual experiment results}
\centering
\begin{tabular}{cc|ccc}
\hline
\multicolumn{2}{c|}{Feature}                                                            & Accuracy         & F1              & RMSE            \\ \hline
\multirow{4}{*}{GloVe}                                                     & + None        & 90.73\%          & 0.9158          & 3.9282          \\
                                                                           & + POS         & 90.73\%          & 0.9122          & 3.8959          \\
                                                                           & + HC          & 92.55\%          & 0.9303          & \textbf{3.3493} \\
                                                                           & + POS + HC    & 93.55\%          & 0.9389          & 3.3650          \\ \hline
\multirow{4}{*}{GPT}                                                       & + None        & 91.55\%          & 0.9224          & 3.7825          \\
                                                                           & + POS         & 92.55\%          & 0.9303          & 4.0275          \\
                                                                           & + HC          & 91.55\%          & 0.9246          & 3.6695          \\
                                                                           & + POS + HC    & 89.82\%          & 0.9076          & 3.7684          \\ \hline
\multirow{4}{*}{RoBERTa}                                                   & + None        & 92.45\%          & 0.9312          & 3.7622          \\
                                                                           & + POS         & 91.64\%          & 0.9231          & 3.8437          \\
                                                                           & + HC          & 93.45\%          & 0.9391          & 3.3852          \\
                                                                           & + POS + HC    & 93.45\%          & 0.9391          & 3.3773          \\ \hline
\multirow{4}{*}{\begin{tabular}[c]{@{}c@{}}Transformer\\ -XL\end{tabular}} & + None        & 92.55\%          & 0.9296          & 4.0078          \\
                                                                           & + POS         & 93.45\%          & 0.9382          & 4.0588          \\
                                                                           & + HC          & \textbf{94.36\%} & \textbf{0.9469} & 3.4866          \\
                                                                           & + POS + HC    & 92.55\%          & 0.9325          & 3.6602          \\ \hline
\end{tabular}
\end{table}

\subsubsection{Unimodal Network}

Table~\ref{tab:unimodal auditory experiment results} is validation results of unimodal networks using acoustic features. The accuracy using VGGish exceeds openSMILE's by over 17\%, which conveys a significant difference in these audio features on performance. Hence, this result establishes a strong point that using an acoustic feature extracted from a pre-trained network outperforms features extracted from scratch.

Textual feature comparing experiment is further conducted by including combinations of using POS and HC features as input. Upon using POS, it is concatenated to the input textual feature to fed into the network. The best performing features for classification and regression are Transformer-XL and GloVe, respectively, according to Table~\ref{tab:unimodal textual experiment results}.

\begin{table}[t]
\setlength{\belowcaptionskip}{-5pt}
\setlength{\abovecaptionskip}{5pt}
\caption{Validation Results of Bimodal Network}
\label{tab:bimodal textual experiment results}
\centering
\begin{tabular}{cc|ccc}
\hline
\multicolumn{2}{c|}{Feature}                                                            & Accuracy         & F1              & RMSE            \\ \hline
\multirow{4}{*}{GloVe}                                                     & + None     & 92.55\%          & 0.9288          & 4.0743          \\
                                                                           & + POS      & 93.55\%          & 0.9398          & 3.7091          \\
                                                                           & + HC       & 90.73\%          & 0.9122          & 3.9138          \\
                                                                           & + POS + HC & 93.55\%          & 0.9382          & 3.4989          \\ \hline
\multirow{4}{*}{GPT}                                                       & + None     & 93.45\%          & 0.9398          & 3.5503          \\
                                                                           & + POS      & 93.45\%          & 0.9398          & 3.9910          \\
                                                                           & + HC       & 91.64\%          & 0.9231          & 3.6334          \\
                                                                           & + POS + HC & 92.55\%          & 0.9318          & 3.5182          \\ \hline
\multirow{4}{*}{RoBERTa}                                                   & + None     & 91.64\%          & 0.9231          & 3.7842          \\
                                                                           & + POS      & 92.55\%          & 0.9318          & 3.6860          \\
                                                                           & + HC       & 93.45\%          & 0.9375          & \textbf{3.4977} \\
                                                                           & + POS + HC & 92.55\%          & 0.9311          & 3.5182          \\ \hline
\multirow{4}{*}{\begin{tabular}[c]{@{}c@{}}Transformer\\ -XL\end{tabular}} & + None     & 91.64\%          & 0.9201          & 4.0703          \\
                                                                           & + POS      & 92.55\%          & 0.9288          & 4.0546          \\
                                                                           & + HC       & 90.73\%          & 0.9114          & 3.7820          \\
                                                                           & + POS + HC & \textbf{94.45\%} & \textbf{0.9454} & 3.6099          \\ \hline
\end{tabular}
\vspace{-4mm}
\end{table}

\subsubsection{Bimodal Network}
We choose VGGish as a fixed auditory input feature for the bimodal network, considering its leading validation performance among other audio features. The use of POS and HC features is performed in the bimodal network as well. Acknowledging the results of unimodal networks, Transformer-XL feature is also well performed in the classification tasks, where RoBERTa feature scores the best root mean squared error (RMSE).

From the observation of this experimental result, it was reasonable to ascertain the best text feature. Notably, using Transformer-XL produced the highest performance in the classification task. Moreover, while comparing the average RMSE scores by feature, RoBERTa outputs the lowest score for both unimodal and bimodal networks. On the other hand, when analyzing performance between additional inputs, only little tendency could be observed. This can be an implication that the quantity of given data may not be sufficient for the additional inputs to exert influence.



\begin{figure}[t]
  \centering
  \includegraphics[width=\linewidth]{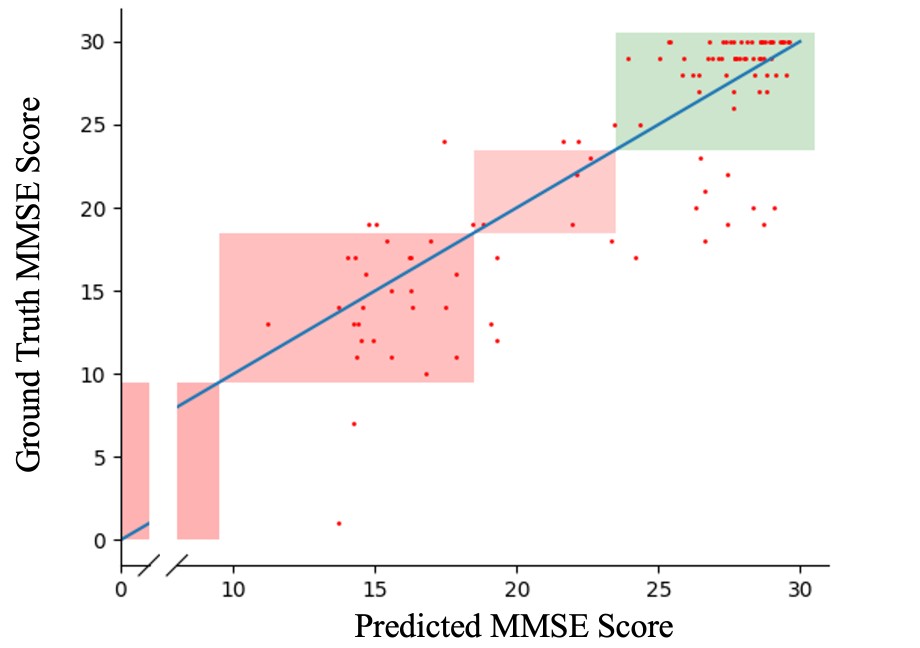}
  \setlength{\abovecaptionskip}{-10pt}
  \setlength{\belowcaptionskip}{-15pt}
  \caption{MMSE comparison graph between regression outputs and ground truth scores - The linear line is a representation of the ideal output with zero error for the task. Each rectangular regions represent dementia severity based on the MMSE score and is shaded respectively as classes of normal, mild, moderate, and severe ranging from high to low values.}
  \label{fig:reg_out}
\end{figure}

\begin{table*}[t]
\setlength{\belowcaptionskip}{-15pt}
\caption{Results of Test Set}
\label{tab:test results}
\centering
\begin{tabular}{c|c|c|cccccc}
\hline
Model                     & Modality                                                                    & Feature                                                                            & Classes & Precision & Recall & F1     & Accuracy                         & RMSE                             \\ \hline
\multirow{2}{*}{Baseline} & \multirow{6}{*}{\begin{tabular}[c]{@{}c@{}}Unimodal\\ Network\end{tabular}} & \multirow{2}{*}{ComParE}                                                           & non-AD  & 0.67      & 0.50   & 0.57   & \multirow{2}{*}{0.625}           & \multirow{2}{*}{6.14}            \\
                          &                                                                             &                                                                                    & AD      & 0.60      & 0.75   & 0.67   &                                  &                                  \\ \cline{1-1} \cline{3-9} 
\multirow{10}{*}{Ours}    &                                                                             & \multirow{2}{*}{VGGish}                                                            & non-AD  & 0.6897    & 0.8333 & 0.7547 & \multirow{2}{*}{0.7292}          & \multirow{2}{*}{5.0765}          \\
                          &                                                                             &                                                                                    & AD      & 0.7895    & 0.6250 & 0.6977 &                                  &                                  \\ \cline{3-9} 
                          &                                                                             & \multirow{2}{*}{Transformer-XL}                                                    & non-AD  & 0.8261    & 0.7917 & 0.8085 & \multirow{2}{*}{\textbf{0.8125}} & \multirow{2}{*}{4.0182}          \\
                          &                                                                             &                                                                                    & AD      & 0.8000    & 0.8333 & 0.8163 &                                  &                                  \\ \cline{2-9} 
                          & \multirow{6}{*}{\begin{tabular}[c]{@{}c@{}}Bimodal\\ Network\end{tabular}}  & \multirow{2}{*}{\begin{tabular}[c]{@{}c@{}}VGGish +\\ GLoVE\end{tabular}}          & non-AD  & 0.7407    & 0.8333 & 0.7843 & \multirow{2}{*}{0.7708}          & \multirow{2}{*}{4.3301}          \\
                          &                                                                             &                                                                                    & AD      & 0.8095    & 0.7083 & 0.7556 &                                  &                                  \\ \cline{3-9} 
                          &                                                                             & \multirow{2}{*}{\begin{tabular}[c]{@{}c@{}}VGGish +\\ Transformer-XL\end{tabular}} & non-AD  & 0.7500    & 0.7500 & 0.7500 & \multirow{2}{*}{0.7500}          & \multirow{2}{*}{\textbf{3.7472}} \\
                          &                                                                             &                                                                                    & AD      & 0.7500    & 0.7500 & 0.7500 &                                  &                                  \\ \cline{3-9} 
                          &                                                                             & \multirow{2}{*}{Ensembled Output}                                                  & non-AD  & 0.7586    & 0.9167 & 0.8302 & \multirow{2}{*}{\textbf{0.8125}} & \multirow{2}{*}{3.7749}          \\
                          &                                                                             &                                                                                    & AD      & 0.8947    & 0.7083 & 0.7907 &                                  &                                  \\ \hline
\end{tabular}
\vspace{-4mm}
\end{table*}

\subsection{Analysis of Regression Task}
Figure~\ref{fig:reg_out} illustrates a graph comparing regression outputs from a bimodal network and the actual patient's corresponding MMSE score during validation stage.
The severity classes, each shaded area in the figure, were categorized based on the MMSE score presented in \cite{crum1993population}.

In this example, VGGish and RoBERTa were used as input features, and the RMSE and r$^2$ value between network outputs and ground truths are 3.5182 and 0.7361, respectively. The output plot shows that the distribution of the network output decreases as the MMSE score decreases, which is inferred to follow the distribution of the given training data. Even though there was no network output below a score of 11, 78.70\% of the points are included in the shaded area. This indicates that classifying severity classes of dementia is possible to some extent, based on regression outputs.


\subsection{Test Set Results}
The test dataset of the ADReSS challenge consists of 48 conversations and can be scored with a total of 5 different submissions. Taking this into account, we use two different models for unimodal and bimodal networks each and an ensembled output of bimodal networks to infer the test data. In the case of the unimodal network, VGGish and Transformer-XL are adopted to represent acoustic and textual modality, respectively. For the bimodal network, GloVe and Transformer-XL are adopted as textual modality regarding on their performance from the validation results. Besides, we select models using POS and HC features along with the bimodality inputs. Lastly, the outputs of the top 5 bimodal networks with high validation results are ensembled and used as the final submission.

The final result for each conversation was deduced by five different models with the same configurations used during the training and validation stage. Combining these results, the final output was concluded using majority voting for AD classification and the median value for the MMSE regression task. The baseline and our test results are presented at Table~\ref{tab:test results}. When using only audio modality, our test accuracy surpasses baseline's by 10\%, where accounting textual modality contributes another 8\% performance gain. Although our textual unimodal model performed the best classification result among the single models, our bimodal network's ensembled output indicates that other bimodal models were able to achieve better performance. Furthermore, the test RMSE implies using both modalities is more advantageous for the regression task.



We could infer from the experimental results that the auditory information led to some performance degradation compared to the textual. This matter can be attributed to the low quality of the audio files provided. In particular, the participant's voice was barely hearable, while it was clear for the investigator's comments in some audio files. Even so, with the methodology to infer AD possible with only recorded audio files, the proposed model can be utilized as a real-world application reflecting on the difficulty of acquiring transcriptions and the target's metadata. The metadata is not dealt with in this work; this is because there was little difference when conditioning age and gender into our model in our prior empirical results.

\section{Conclusion}


This paper demonstrates extracted features from pre-trained networks are satisfactory for handling small amounts of data, to recognize Alzheimer's Dementia. The proposed model can compute variable lengths of dialogue and also introduce productive methods to fit the network with a little amount of data. Furthermore, our model does not require any metadata and also can perform well without transcript, which may be practical in real-world situations. Our test result outperforms baseline's with both tasks, and our regression results imply the potential of network classifying classes of cognitive impairment based on the MMSE score.





For future work, with the expectation of performance gain, mechanisms effectively fusioning different modality features \cite{hori2017attention} \cite{yilmaz2012non} can be applied in the model architecture.

\section{Acknowledgements}
This work was supported partly by Institute for Information \& Communications Technology Planning \& Evaluation(IITP) grant funded by the Korea government(MSIT) (No.2019-0-01367, BabyMind) and partly by Next-Generation Information Computing Development Program through the National Research Foundation of Korea(NRF) funded by the Ministry of Science and ICT(NRF-2017M3C4A7078548). Also, the authors would like to thank Jeonghyun Yoon, and Ayoung Choi for their fruitful comments, and inspiration.

\bibliographystyle{IEEEtran}

\bibliography{mybib.bib}

\clearpage

\onecolumn

\setcounter{section}{0}
\renewcommand*{\thesection}{\Alph{section}}
\section{Full Test Results}

\begin{center}

We disclose the full test results in hopes of providing some insights to researchers in this field.

\end{center}

\begin{table}[hbt!]
\setlength{\belowcaptionskip}{30pt}
\caption{Full Test Results of Textual Unimodal Network}
\label{tab:full test results of textual unimodal network}
\centering
\begin{tabular}{cc|ccc}
\hline
\multicolumn{2}{c|}{Feature}                 & Accuracy         & F1     & RMSE            \\ \hline
Linguistics                     & (Baseline) & 75.00\%          & 0.7450 & 5.20            \\ \hline
\multirow{4}{*}{GloVe}          & + None     & 77.08\%          & 0.7442 & 4.8110          \\
                                & + POS      & 70.83\%          & 0.6111 & 4.1608          \\
                                & + HC       & 79.17\%          & 0.7826 & 4.4628          \\
                                & + POS + HC & 77.08\%          & 0.7317 & 3.8891          \\ \hline
\multirow{4}{*}{GPT}            & + None     & 79.17\%          & 0.7727 & 4.4253          \\
                                & + POS      & 72.92\%          & 0.7111 & 5.1174          \\
                                & + HC       & 81.25\%          & 0.8000 & 4.1433          \\
                                & + POS + HC & 83.33\%          & 0.8182 & \textbf{3.5237} \\ \hline
\multirow{4}{*}{RoBERTa}        & + None     & 77.08\%          & 0.7179 & 4.3253          \\
                                & + POS      & 85.42\%          & 0.8372 & 4.3157          \\
                                & + HC       & 81.25\%          & 0.8085 & 3.5824          \\
                                & + POS + HC & 81.25\%          & 0.7907 & 3.7137          \\ \hline
\multirow{4}{*}{Transformer-XL} & + None     & 83.33\%          & 0.8400 & 4.4371          \\
                                & + POS      & 75.00\%          & 0.7391 & 4.6075          \\
                                & + HC       & 81.25\%          & 0.8163 & 4.0182          \\
                                & + POS + HC & \textbf{89.58\%} & \textbf{0.8889} & 4.2254          \\ \hline
\end{tabular}
\end{table}

\begin{table}[hbt!]
\setlength{\belowcaptionskip}{30pt}
\caption{Full Test Results of Bimodal Network}
\label{tab:full test results of bimodal network}
\centering
\begin{tabular}{cc|ccc}
\hline
\multicolumn{2}{c|}{Feature}                 & Accuracy         & F1     & RMSE            \\ \hline
\multirow{4}{*}{GloVe}          & + None     & 66.67\%          & 0.6364 & 4.8045          \\
                                & + POS      & 72.92\%          & 0.6667 & 4.5939          \\
                                & + HC       & 72.92\%          & 0.7111 & 4.5712          \\
                                & + POS + HC & 77.08\%          & 0.7556 & 4.3301          \\ \hline
\multirow{4}{*}{GPT}            & + None     & 79.17\%          & 0.7619 & 4.5092          \\
                                & + POS      & 77.08\%          & 0.7442 & 3.9396          \\
                                & + HC       & 75.00\%          & 0.7273 & 3.8837          \\
                                & + POS + HC & 77.08\%          & 0.7317 & 4.1408          \\ \hline
\multirow{4}{*}{RoBERTa}        & + None     & 81.25\%          & 0.7692 & 4.4721          \\
                                & + POS      & \textbf{85.42\%} & \textbf{0.8293} & 4.0415          \\
                                & + HC       & 83.33\%          & 0.8182 & 4.1508          \\
                                & + POS + HC & 79.17\%          & 0.7619 & \textbf{3.5178} \\ \hline
\multirow{4}{*}{Transformer-XL} & + None     & 79.17\%          & 0.7727 & 4.7500          \\
                                & + POS      & 70.83\%          & 0.6667 & 4.5116          \\
                                & + HC       & 75.00\%          & 0.7391 & 3.8052          \\
                                & + POS + HC & 75.00\%          & 0.7500 & 3.7472          \\ \hline
\end{tabular}
\end{table}

\end{document}